\documentclass[conference]{IEEEtran}
\IEEEoverridecommandlockouts

\usepackage{amsmath,amsfonts}
\usepackage{graphicx}
\usepackage{textcomp}
\usepackage{xcolor}
\usepackage{stfloats}
\usepackage{tikz}
\usepackage{pgfplots}
\pgfplotsset{compat=1.18} % Ensure compatibility with your TikZ installation
\usepackage{adjustbox}
\usepackage{subcaption}

\usepackage{fancyhdr}
\usepackage{listings}
\usepackage{xcolor}
\usepackage{booktabs} 
\usepackage{threeparttable}
\usepackage{multirow}
\usepackage[linesnumbered,ruled,vlined]{algorithm2e}
\usepackage{url}
\usepackage{multicol}
\usepackage{tcolorbox}
\usepackage{enumitem}
\usepackage{array}     
\usepackage{graphicx} 
\usepackage{arydshln}
\usepackage{makecell} 
\tcbuselibrary{listingsutf8}
\usepackage{pifont}
\usepackage[hidelinks]{hyperref}
\usepackage{bigdelim}

\newenvironment{packeditemize}{
	\begin{list}{$\bullet$}{
			\setlength{\labelwidth}{4pt}
			\setlength{\itemsep}{0pt}
			\setlength{\leftmargin}{\labelwidth}
			\addtolength{\leftmargin}{\labelsep}
			\setlength{\parindent}{0pt}
			\setlength{\listparindent}{\parindent}
			\setlength{\parsep}{0pt}
			\setlength{\topsep}{1pt}}}{\end{list}}

\begin{document}

\title{Large Language Models for Cryptocurrency Transaction Analysis: A Bitcoin Case Study
\thanks{Yuchen Lei and Yuexin Xiang contributed equally to this work.}
\thanks{*Yuexin Xiang is the corresponding author.}
}

\author{
Yuchen Lei$^{1}$,
Yuexin Xiang$^{2,*}$,
Tsz Hon Yuen$^{2}$,
Andreas Deppeler$^{3}$,
Qin Wang$^{4}$,
Kim-Kwang Raymond Choo$^{5}$
\\[1ex]
\IEEEauthorblockA{$^{1}$ Independent Researcher}
\IEEEauthorblockA{$^{2}$ Faculty of Information Technology, Monash University, Melbourne, Australia}
\IEEEauthorblockA{$^{3}$ School of Business, Monash University Malaysia, Selangor, Malaysia}
\IEEEauthorblockA{$^{4}$ CSIRO, Sydney, Australia}
\IEEEauthorblockA{$^{5}$ Department of Information Systems and Cybersecurity, University of Texas at San Antonio, TX, USA}
}

% \author{
% Yuchen Lei$^{1}$,
% Yuexin Xiang$^{2}$,
% Tsz Hon Yuen$^{2}$,
% Andreas Deppeler$^{3}$,
% Qin Wang$^{4}$,
% Kim-Kwang Raymond Choo$^{5}$
% \\[1ex]
% \IEEEauthorblockA{$^{1}$ Independent Researcher}
% \IEEEauthorblockA{$^{2}$ Faculty of Information Technology, Monash University, Melbourne, Australia}
% \IEEEauthorblockA{$^{3}$ School of Business, Monash University Malaysia, Selangor, Malaysia}
% \IEEEauthorblockA{$^{4}$ CSIRO, Sydney, Australia}
% \IEEEauthorblockA{$^{5}$ Department of Information Systems and Cybersecurity, University of Texas at San Antonio, TX, USA}
% }

% \author{
%   \IEEEauthorblockN{Anonymous Authors}
%   \IEEEauthorblockA{Anonymous Affiliations}
% }

\maketitle

\begin{abstract}
Cryptocurrency transaction analysis often relies on black-box machine learning models that achieve high predictive performance but offer limited interpretability of transaction behaviors. This paper investigates the capabilities of large language models (LLMs) in analyzing real-world Bitcoin transaction graphs through a three-level framework: foundational metrics, characteristic overview, and contextual interpretation. To address token constraints, we introduce LLM4TG, a compact, human-readable graph representation, and CETraS, a connectivity-enhanced graph sampling algorithm. Experiments across five LLMs show that node-level information extraction achieves over 98.50\% accuracy on most metrics, while meaningful graph characteristics are identified in up to 95.00\% of responses. For address classification with limited labeled examples, top-1 accuracy reaches 50.49\% using raw graphs, while top-3 accuracy reaches 72.43\% using engineered features. These findings demonstrate the potential of LLMs for transaction graph analysis while revealing limitations in global graph reasoning, precise address classification, and explanation reliability.
\end{abstract}


%==========================================
\section{Introduction}
%==========================================

Cryptocurrency transaction analysis plays an important role in understanding address behaviors and identifying illicit activities such as scams and money laundering. Existing approaches often rely on engineered features and specialized machine learning models, which can achieve effective classification but provide limited interpretability. Large language models (LLMs) offer an alternative through their ability to process structured information and generate natural-language explanations.

Recent studies have explored LLMs for graph analysis, including synthetic graphs, citation networks, and knowledge graphs~\cite{guo2023gpt4graph,hu2023beyond,wang2023can,zhang2023llm4dyg}. However, their capabilities in analyzing real-world cryptocurrency transaction graphs remain insufficiently understood. Unlike conventional graph datasets, Bitcoin transaction graphs represent flows between address and transaction nodes, with attributes such as transaction values, degrees, and timestamps~\cite{10375557}. Their size and structural complexity also create challenges for LLMs, particularly under input token constraints.

To investigate these challenges, we address three research questions:
\begin{itemize}
    \item \textbf{RQ1:} What graph representation formats are effective for LLM-based Bitcoin transaction analysis?
    \item \textbf{RQ2:} How can we evaluate LLMs' understanding of Bitcoin transaction graphs?
    \item \textbf{RQ3:} How does using engineered graph features differ from using raw transaction graphs?
\end{itemize}

For RQ1, we introduce LLM4TG, a compact, human-readable graph representation designed to reduce token consumption, and CETraS, a connectivity-enhanced sampling method that preserves important transaction information. For RQ2, we establish a three-level evaluation framework: \emph{foundational metrics}, assessing the extraction of node attributes and graph statistics; \emph{characteristic overview}, examining the identification of distinctive structural and transactional patterns; and \emph{contextual interpretation}, evaluating address-type classification with explanations. For RQ3, we compare LLM performance using raw transaction graphs and engineered graph features.

Our main contributions are summarized as follows:
\begin{itemize}
    \item We present a three-level framework for systematically evaluating LLMs' capabilities in Bitcoin transaction graph analysis, covering local information extraction, graph characterization, and contextual interpretation.
    \item We develop LLM4TG and CETraS to improve the token efficiency of transaction graph representations while retaining transaction-relevant information and connectivity.
    \item We conduct quantitative and qualitative evaluations across five LLMs from the GPT, DeepSeek, and LLaMA families. Our results reveal strong performance in extracting local information and identifying graph characteristics, but substantially weaker performance in global reasoning and exact address classification. We further examine the differences between raw-graph and feature-based inputs.
\end{itemize}

%=========================================================
\section{Related Work}
%=========================================================

\noindent\textbf{Cryptocurrency Transaction Analysis.}
Previous studies have investigated cryptocurrency ecosystems and address behaviors through empirical analysis~\cite{wang2024cryptocurrency,huang2023miracle,wang2022empirical,chen2020understanding,gao2020tracking}, including Bitcoin mining pools~\cite{tovanich2021empirical,hou2020study}, cryptocurrency scams~\cite{li2023double,he2023txphishscope}, cybercrime networks~\cite{gomez2022watch}, money laundering~\cite{10371347}, and exchange trading activities~\cite{chen2022cryptocurrency}. Graph-based approaches further characterize cryptocurrency transaction networks using complex network analysis~\cite{nerurkar2021dissecting,serena2022cryptocurrencies,tao2021complex,guo2019graph} and identify suspicious fund flows~\cite{lin2024denseflow}. Machine learning methods have been applied to illicit activity detection and address classification~\cite{chaudhari2021towards,rathore2022mixers,wahrstatter2023improving}, supported by benchmark datasets~\cite{weber2019anti,10375557,xiang2022leveraging} and advanced architectures such as BlockGPT~\cite{gai2023blockchain}.

\smallskip
\noindent\textbf{LLMs in Graph Analysis.}
Recent studies have explored LLMs for graph reasoning, classification, and GNN-related tasks~\cite{li2023survey,wang2023can,yu2023predicting,tang2025grapharena,hu2023beyond,liu2023evaluating}. Other work investigates LLMs' ability to process structured data through prompting~\cite{sui2023gpt4table,jiang2023structgpt}, integrate graph information for downstream tasks~\cite{das2023modality,chen2023exploring,chen2023labelf,guo2023gpt4graph}, and recall factual knowledge~\cite{sun2023head}. These studies reveal both the potential and limitations of LLMs in structured reasoning.

\smallskip
\noindent\textbf{Research Gap.}
Existing LLM-based graph studies primarily focus on synthetic graphs, knowledge graphs, and general graph-learning tasks. Their findings provide limited evidence of how LLMs analyze real-world cryptocurrency transaction graphs, particularly under token constraints. Moreover, the effects of graph representation, sampling, and engineered features on LLM-based transaction analysis remain insufficiently examined. This work addresses these gaps through a three-level evaluation framework, LLM4TG, and CETraS.

\begin{figure}[!b]
  \centering
  \includegraphics[width=0.8\linewidth]{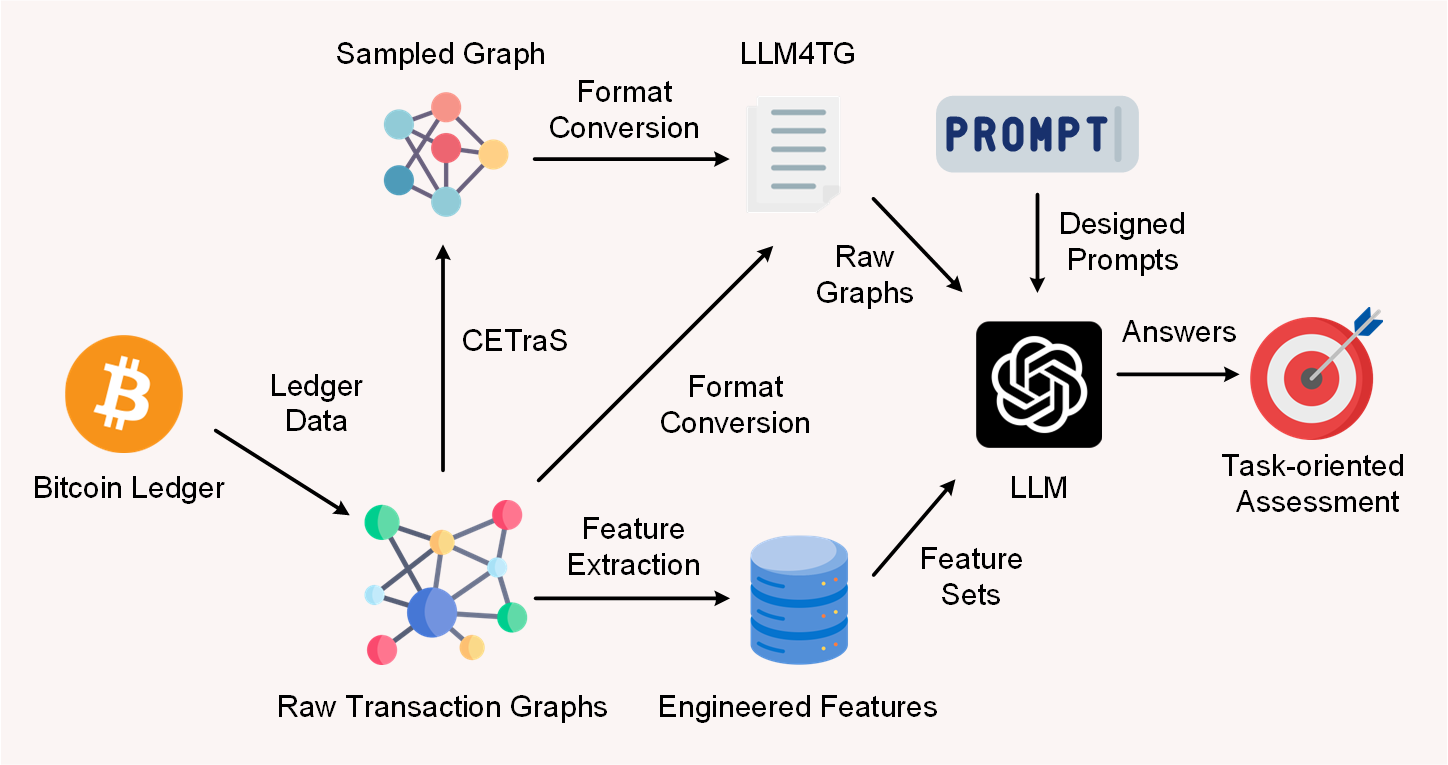}
  \caption{LLM evaluation framework for Bitcoin transaction}
  \label{fw}
\end{figure}

%=========================================================
\section{Methodology}
\label{sec-method}
%=========================================================

\subsection{Framework Overview}

We present our framework in Fig.~\ref{fw}. The first step is to construct the Bitcoin transaction graph using historical on-chain data. The second step is to extract corresponding subgraphs (or subgraphs sampled by CETraS) and graph features for labeled Bitcoin addresses. Finally, raw graphs formatted in LLM4TG and graph features are provided as inputs to LLMs with task-specific prompts for different levels of understanding.

\subsection{LLM4TG}
\label{subsec-llm4tg}

We introduce LLM4TG, a new graph representation format designed to optimize the analysis of transaction graphs using LLMs. This format is text-based and human-readable, minimizing syntactic noise/redundancy while reducing token usage and preserving data integrity.

In our approach, LLM4TG captures essential node information and integrates edge details directly within the nodes. It organizes nodes into layers based on their type, either address or transaction, thereby maintaining the structural integrity of the graph. This hierarchical layering provides a segmented and clear overview of the network's dynamics. Each layer categorizes nodes which are further defined by properties such as degrees and token amounts, simplifying the analysis and enhancing readability.

We denote \texttt{T} as transaction and \texttt{A} as address. \texttt{<NodeID>} represents the node's ID, \texttt{<Number>} represents an integer. \texttt{<Float>} represents a real number. The syntax is displayed in Listing~\ref{lst1}.

This format provides three key advantages for representing and analyzing transaction graphs: 1) It organizes nodes into type-specific layers, mirroring the structure of original transaction graphs; 2) It efficiently utilizes the limited token budget of LLMs by allowing more data to be encoded; 3) It improves the interpretability of graph data for LLMs by organizing node attributes into closely associated key-value pairs.

To further demonstrate the effectiveness of LLM4TG, we compared its token consumption with other formats for the same graphs, as shown in Fig. \ref{fig:tokens}. This comparison reveals that LLM4TG experiences a more gradual increase in token usage and consistently stays within the GPT-4/4o token limit across various graph sizes, making it a more efficient format, especially for larger graphs. Further details and discussion are presented in Section~\ref{sec-discu}.

\smallskip
\begin{lstlisting}  
<LLM4TG> ::= <GraphLayer>+
<GraphLayer> ::= "Layer" <Number> ":" <NodeCount> <NodeType> "nodes" <NewLine> <Node>+
<NodeCount> ::= <Number>
<NodeType> ::= "address" | "transaction"
<Node> ::= <NodeA> | <NodeT>
<NodeA> ::= <NodeID> "address" ":" <PropertiesA> <NewLine>
<NodeT> ::= <NodeID> "transaction" ":" <PropertiesT> <NewLine>
<PropertiesA> ::= "{" <PropertyA> ("," <PropertyA>)* "}"
<PropertiesT> ::= "{" <PropertyT> ("," <PropertyT>)* "}"
<Property> ::= <InDegree> | <OutDegree> | <InValue> | <OutValue>
<PropertyA> ::= <Property> | <TimeRange>
<PropertyT> ::= <Property> | <InNodes> | <OutNodes>
<InDegree> ::= "in_degree:" <Number>
<OutDegree> ::= "out_degree:" <Number>
<InValue> ::= "in_value:" <Float>
<OutValue> ::= "out_value:" <Float>
<TimeRange> ::= "time_range:" <Number>
<InNodes> ::= "in_nodes:" "[" <NodeIDList> "]"
<OutNodes> ::= "out_nodes:" "[" <NodeIDList> "]"
<NodeIDList> ::= <NodeID> ("," <NodeID>)* | <Empty>
\end{lstlisting}

\subsection{CETraS}
\label{subsec-cetras}

Despite LLM4TG's efficiency, some transaction graphs are too large for tasks like classification that involve few-shot learning, which processes multiple graphs at once. To tackle this, we introduce CETraS, a method that condenses mid-sized transaction graphs while preserving important nodes and their connectivity.

CETraS assigns an importance score $I_{\mathrm{node}}$ to each node based on its transaction amount, degree, and distance from the root node $n_0$. Specifically, $a_{\mathrm{in}}$ and $a_{\mathrm{out}}$ denote the input and output token amounts, while $d_{\mathrm{in}}$ and $d_{\mathrm{out}}$ represent the in-degree and out-degree. $L_s$ is the shortest-path distance from the node to $n_0$, computed on the undirected graph. The parameter $\beta$ controls the relative importance of node degree. We set $\beta=2$ to prioritize connectivity. The importance score is calculated as:

\begin{equation}
I_{\mathrm{node}} =
\frac{
\log(a_{\mathrm{in}}+a_{\mathrm{out}}+1)
+\beta\log(d_{\mathrm{in}}+d_{\mathrm{out}}+1)
}{
L_s+1
}.
\end{equation}

As detailed in Algorithm~\ref{alg:cetras}, CETraS prioritizes nodes with lower importance for removal. Specifically, it assigns removal probabilities inversely proportional to node importance and samples nodes for removal without replacement. The root node $n_0$ is always retained, and nodes unreachable from the root are excluded. Given a target node count $N_{\mathrm{target}}$, the algorithm initially retains $\min(N_{\mathrm{target}},|V(G)|)$ nodes.

To preserve connectivity, CETraS subsequently restores the shortest paths from the root node to all initially retained nodes. The final sampled graph is induced by the resulting node set, preserving the original edges between retained nodes. Consequently, the final graph may contain more than $N_{\mathrm{target}}$ nodes due to path restoration. Node attributes are computed from the original graph, while the number of nodes in each layer is recalculated after sampling to ensure consistency with the resulting LLM4TG representation.

Unlike existing graph summarization methods~\cite{hajiabadi2021graph,kang2022personalized,lee2020ssumm,lai2023optimized} that primarily focus on preserving the accuracy of structural queries or computations on large-scale graphs, CETraS focuses on retaining transaction-relevant information and connectivity in moderately large transaction graphs for LLM-based analysis.

\begin{algorithm}[t]
\caption{CETraS}
\label{alg:cetras}
\footnotesize
\KwIn{Original transaction graph $G$; Target number of nodes to retain $N_{\mathrm{target}}$}
\KwOut{Sampled transaction graph $G_{\mathrm{sampled}}$}

\SetKwFunction{SampleGraph}{SampleGraph}
\SetKwProg{Fn}{Function}{:}{end}
\Fn{\SampleGraph{$G,N_{\mathrm{target}}$}}{
    $I_{\mathrm{node}} \gets \{v:I_{\mathrm{node}}[v] \text{ for each } v \in V(G)\}$\;
    $P_{\mathrm{node}} \gets 1/I_{\mathrm{node}}$\;
    $P_{n_0} \gets 0$\;
    $P_{\mathrm{sum}} \gets \sum P_{\mathrm{node}}$\;
    \ForEach{node $n \in G$}{
        $P_n \gets P_n/P_{\mathrm{sum}}$\;
    }
    $N_{\mathrm{remove}} \gets |V(G)|-\min\{N_{\mathrm{target}},|V(G)|\}$\;
    $G_{\mathrm{removed}} \gets$ Sample $N_{\mathrm{remove}}$ nodes
    without replacement using $P_{\mathrm{node}}$\;
    $G_{\mathrm{subset}} \gets V(G)\setminus G_{\mathrm{removed}}$\;
    $G_{\mathrm{kept}} \gets G_{\mathrm{subset}}$\;

    \ForEach{node $n \in G_{\mathrm{subset}}$}{
        $p \gets$ Compute shortest path from $n_0$ to $n$ in $G$
        (undirected)\;
        \ForEach{node $m \in p$}{
            Add node $m$ to $G_{\mathrm{kept}}$\;
        }
    }
    $G_{\mathrm{sampled}} \gets G[G_{\mathrm{kept}}]$\;
    \ForEach{layer $l$ in $G_{\mathrm{sampled}}$}{
        $C_l \gets |\{n\in G_{\mathrm{kept}}:L_s(n)=l\}|$\;
    }
    \Return{$G_{\mathrm{sampled}}$}\;
}
\end{algorithm}

%============================================================
\section{Evaluation and Analysis}
\label{sec:4ex}
%============================================================

\subsection{Settings}

\noindent\textbf{Dataset.} We use two datasets for experiments, both constructed from the whole Bitcoin transaction graph, which spans a 22-month period (12 July 2019 – 26 May 2021). Specifically, BASD~\cite{xiang2022leveraging} includes eight types of subgraphs, each starting from a labeled address (denoted as $n_0$) and extending up to five hops with at most 3,000 nodes. It is one of the few datasets that provides structured graph-level representations of Bitcoin addresses. Meanwhile, BABD~\cite{10375557} contains labeled Bitcoin addresses, each associated with 148 engineered features. It is one of the largest publicly available datasets for Bitcoin address behavior analysis.

\smallskip
\noindent\textbf{LLM Selection.} We select five LLMs for evaluation: GPT-3.5 (gpt-3.5-turbo), GPT-4 (gpt-4-turbo), GPT-4o (gpt-4o), DeepSeek (deepseekv3), and LLaMA (llama3.3-70B). These models are chosen for their strong overall performance and representativeness in specific tasks. Due to differences in their input length capacities (e.g., up to 8,192 tokens for llama3.3-70B), not all models are evaluated on every task. All models are accessed via application programming interfaces (APIs), which support prompt-based querying and response generation.

\smallskip
\noindent\textbf{Graph Formatting and Model Allocation.} We format raw transaction graphs as LLM4TG in the experiments. We use CETraS in level 2 and level 3 due to the token limit in few-shot prompts. We apply GPT-3.5 only in level 3 feature-based classification, given its particularly tight token limits (see Fig.\ref{fig:tokens}).  Experiments are conducted on BASD dataset subsets and its corresponding BABD addresses. 

\smallskip
\noindent\textbf{Prompt Engineering and Design.} We employ few-shot prompting~\cite{wei2022chain} to interact with LLMs, providing only a small number of examples to guide task execution. To further align prompts with the model’s capabilities and preferences, we use an LLM-feedback generation strategy, wherein the model is queried with task-specific requirements and subsequently designs the corresponding system prompts. All code and data are available at  GitHub.\footnote{\url{https://github.com/yuchen-lei/llm4tg}}

\smallskip
\noindent\textbf{Experimental Setup.}
All experiments are conducted on a workstation equipped with an AMD Ryzen Threadripper PRO 5995WX and 256 GB of RAM, running Ubuntu 22.04.

%-------------------------
\subsection{Level 1 - Foundational Metrics}
\label{s-lv-1}
%-------------------------

We evaluate LLMs on sampled transaction graphs with selected nodes across layers. The evaluation adopts twelve metrics covering three perspectives: response metrics, global metrics, and node metrics:

\begin{itemize}
\item\textbf{Response Metrics.} To evaluate if LLM responses are correctly structured, \emph{struct\_correctness} is applied.
\item\textbf{Global Metrics.} To assess the ability of LLMs to basic metrics understanding for the entire transaction graphs, LLMs need to find the node with the largest in/out-degree (\emph{global\_in/out\_degree}), the node with the largest in/out-value (\emph{global\_in/out\_value}), and the node with the largest difference between input and output values/degrees (\emph{global\_diff\_degree/value}).
\item\textbf{Node Metrics.} To investigate the capability of LLMs to understand foundational information of concrete nodes in transaction graphs, LLMs need to obtain the node's in/out-degree (\emph{node\_in/out\_degree}), the node's in/out-value (\emph{node\_in/out\_value}), and the node's special information (\emph{node\_special\_info\_a/t}). The special information for the address node is time interval; while for the transaction node is if a specific node exists in the input/output node sets.
\end{itemize}

\begin{table}[!htbp]
\centering
\footnotesize
\caption{LLM Capability on foundational metrics}
\label{lv1}
\renewcommand{\arraystretch}{1}
\setlength{\tabcolsep}{12pt} 
\resizebox{0.8\linewidth}{!}{
\begin{tabular}{l cc}
\toprule
\textbf{Metrics} & \textbf{GPT-4} & \textbf{GPT-4o}  \\
\midrule
\emph{struct\_correctness} & 80.00\%  & 100.00\%  \\
\midrule
\emph{global\_in\_degree} & 50.00\%  & 44.00\% \\
\emph{global\_out\_degree} & 50.00\% & 58.00\%  \\
\emph{global\_in\_value} & 37.50\% & 56.00\% \\
\emph{global\_out\_value} & 35.00\% & 48.00\% \\
\emph{global\_diff\_degree} & 27.50\% & 34.00\% \\
\emph{global\_diff\_value} & 27.50\% & 24.00\% \\
\midrule
\emph{node\_in\_degree} & 99.25\% & 99.40\% \\
\emph{node\_out\_degree} & 100.00\% & 99.80\% \\
\emph{node\_in\_value} & 98.50\% & 99.00\% \\
\emph{node\_out\_value} & 98.50\% & 99.00\%  \\
\emph{node\_special\_info\_a} & 99.08\% & 100.00\% \\
\emph{node\_special\_info\_t} & 70.88\% & 67.63\%  \\
\bottomrule
\end{tabular}
}
\end{table}

\smallskip
\noindent\textbf{Results Analysis.} Table \ref{lv1} demonstrates that LLMs are excellent at node metrics. Accuracy for most metrics is between 98.50\% and 100.00\%. \emph{node\_special\_info\_t} is the exception. This may be due to the limited capability of LLMs to match many structurally similar data in transaction graphs. Compared with node metrics, however, for the global metrics, the accuracy significantly drops, ranging from 24.00\% to 58.00\%. Especially, we find that compared with the other metrics (35.00\% to 58.00\%) in global metrics, difference-related metrics are relatively low (24.00\% to 34.00\%). The reason for this may be that the capability of LLMs to calculate or compare is limited.

% Perspective of different LLMs
For different LLMs, i.e., GPT-4 and GPT4o, the most various points are \emph{struct\_correctness} in response metrics and \emph{global\_in/out\_value} in global metrics. The enhancement of \emph{struct\_correctness} represents that the update and optimization of the LLMs improve the quality of the response format, which completely follows the requirements in prompts. Likewise, slightly improved \emph{global\_in/out\_value} also illustrate the effectiveness of model upgrade. However, most of the metrics remain at similar levels, which shows the inherent flaws of LLMs for basic information understanding in transaction graphs, especially global metrics.

\vspace{0.5\baselineskip}
\begin{center}
  % \fbox{
   \colorbox{blue!7}{
         \begin{minipage}{0.89\linewidth}
            \textbf{Level 1 Findings.}
            LLMs accurately extract explicit node-level information but perform substantially worse on global metrics, particularly cross-node numerical comparisons. This reveals limitations in holistic graph reasoning despite strong local information retrieval.
         \end{minipage}
      }
 %  }
\end{center}

%-------------------------
\subsection{Level 2 - Characteristic Overview}
\label{s-lv-2}
%-------------------------

To evaluate LLMs’ ability to identify distinctive subgraph characteristics, we provide representative reference subgraphs and test samples without labels to avoid bias. We divide the quality of LLMs' responses into three quality levels:
\begin{itemize}
\item \emph{High-quality} responses exclude irrelevant, flawed, or incorrect information.
\item \emph{Average-quality} responses include irrelevant or flawed but no incorrect information. 
\item \emph{Low-quality} responses contain incorrect information. 
\end{itemize}
In this context, irrelevant responses are correct but do not answer the question or provide useful information;  Flawed responses have small mistakes or misleading parts but are mostly right; Incorrect responses contain false or completely wrong information.

\begin{table}[!htbp]
\centering
\footnotesize
\caption{LLM capability on characteristic overview}
\label{tab:quality-gpt4-gpt4o}
\renewcommand{\arraystretch}{1.20}
\setlength{\tabcolsep}{6pt}
\begin{tabular}{lc@{\hspace{2.5em}}cc}
\toprule
\textbf{Metric} & & \textbf{GPT-4} & \textbf{GPT-4o} \\
\midrule
High-quality & & \textbf{62.50\%} & \textbf{82.50\%} \\
\ldelim\{{3}{1.2em}[Avg-quality] & Total     & 26.25\% & 13.75\% \\
                                  & Flawed              & \underline{\textbf{7.50\%}}  & \underline{\textbf{12.50\%}} \\
                                  & Irrelevant          & \underline{18.75\%} & \underline{1.25\%}  \\
Low-quality & & 11.25\% & 3.75\% \\
\midrule
Meaningful  & & \textbf{70.00\%} & \textbf{95.00\%} \\
\bottomrule
\end{tabular}
\end{table}

\smallskip
\noindent\textbf{Overall Results.} Table~\ref{tab:quality-gpt4-gpt4o} shows that GPT-4o delivers substantially higher response quality than GPT-4. High-quality responses make up 82.50\% of GPT-4o’s responses, compared with 62.50\% for GPT-4. For average-quality responses, GPT-4o scores 13.75\%, with 12.50\% flawed and 1.25\% irrelevant; While GPT-4 records 26.25\%, with 7.50\% flawed and 18.75\% irrelevant. Low-quality responses are 3.75\% for GPT-4o and 11.25\% for GPT-4. Overall, GPT-4o produces meaningful outputs in 95.00\% of cases versus 70.00\% for GPT-4.

\smallskip
\noindent\textbf{Illustrative Examples.} We select two examples \texttt{bc1qah\footnote{\texttt{bc1qahe54yxl33clnwdtleuh4cw0fw4df62t0tnuk2}}} and \texttt{bc1q4w\footnote{\texttt{bc1q4w090gzj7m9hy39l8e48gzz4uvdfn2x9y9g70h}}}, to ensure each category is included to further illustrate our results, combined with visualization by \emph{Gephi}. We apply the address identifier of node $n_0$ to represent the corresponding subgraph.

\begin{figure}[!]
\begin{subfigure}[t]{0.45\linewidth}
\includegraphics[width=\linewidth]{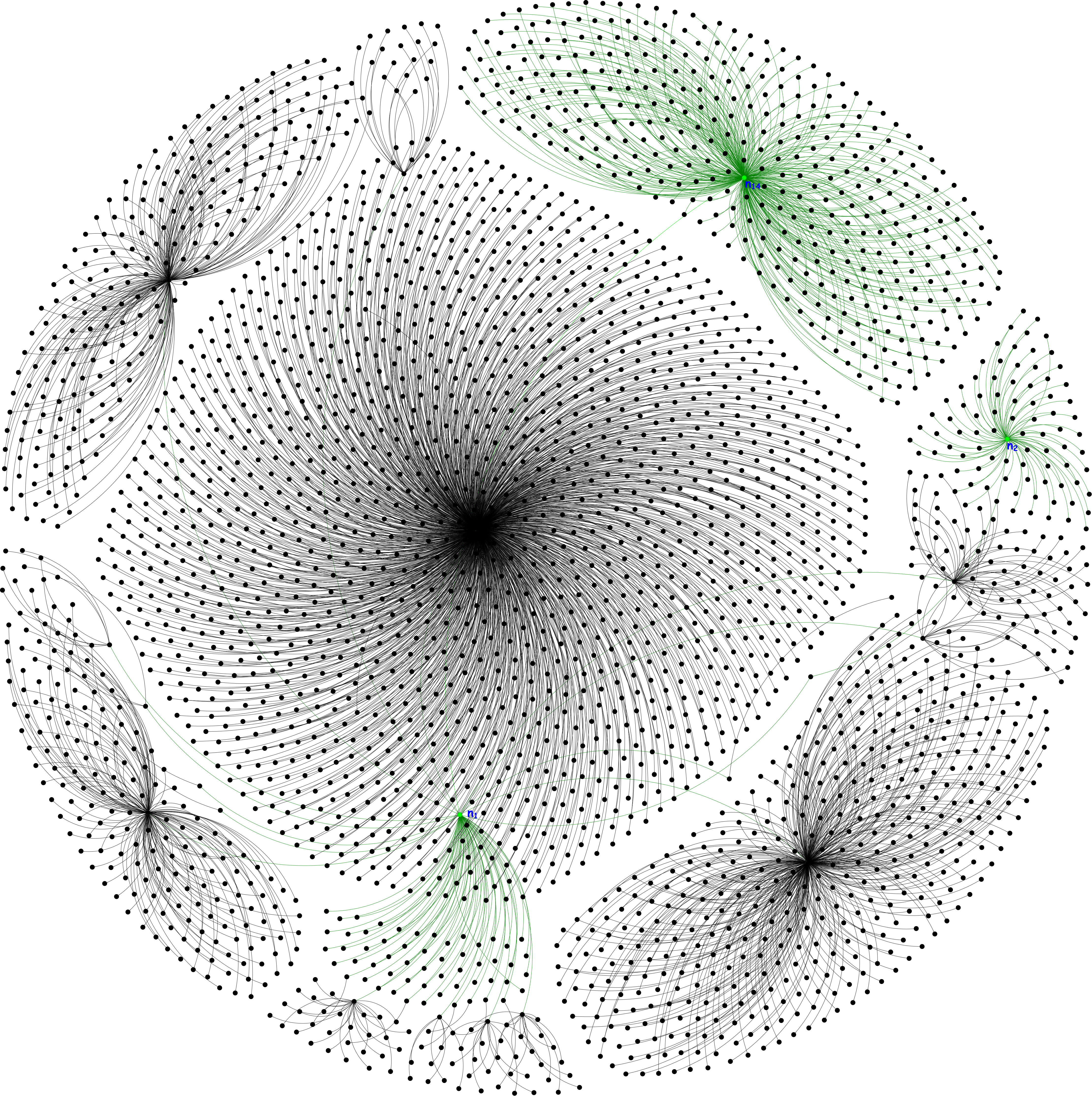}
\caption{\texttt{bc1qah} features high out-degrees for $n_1$, $n_2$, and both high in-/out-degree for $n_{14}$.}
\label{lv2-1}
\end{subfigure}
\hfill
\begin{subfigure}[t]{0.45\linewidth}
\includegraphics[width=\linewidth]{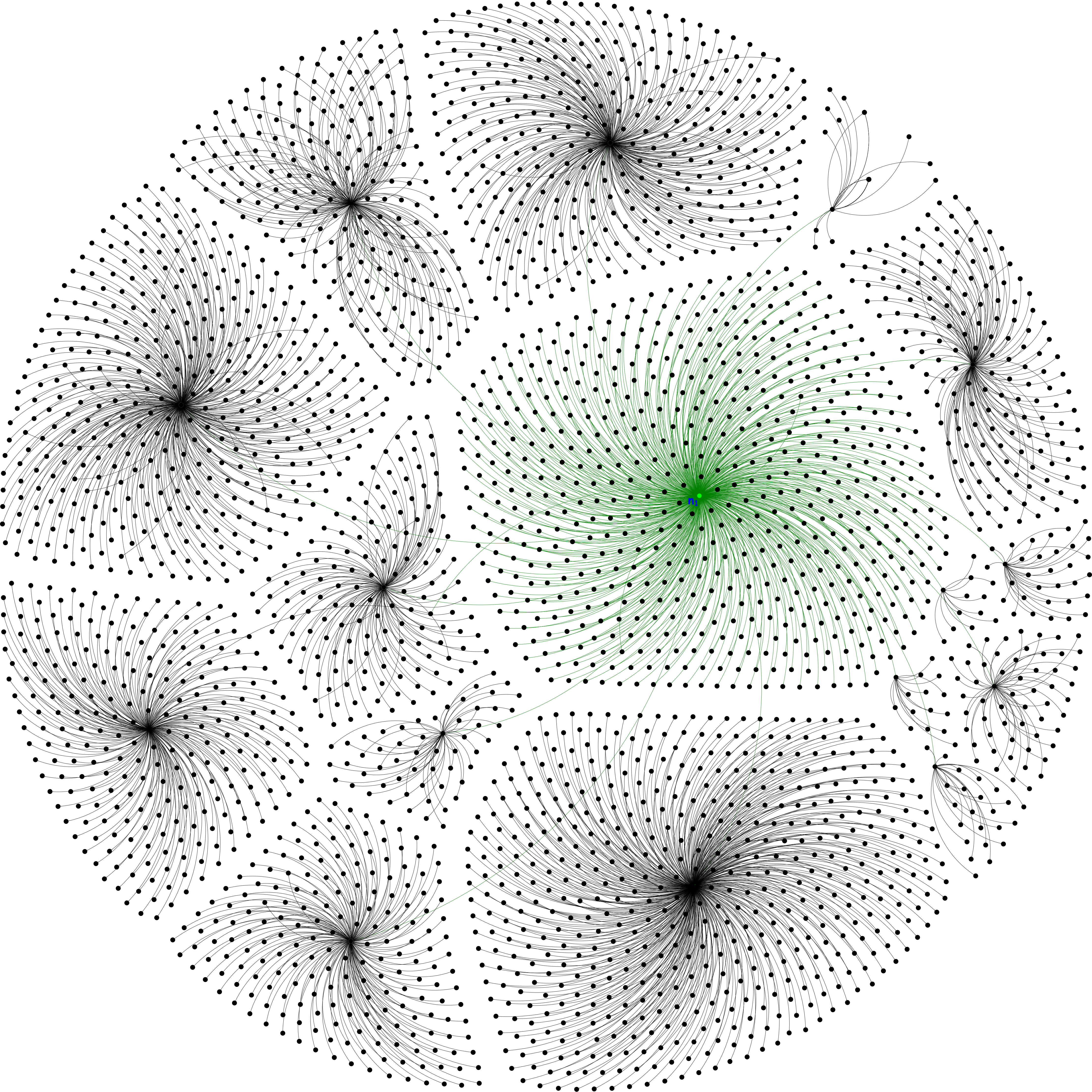}
\vspace{-0.15in}
\caption{In constract, \texttt{bc1q4w} only features high out-degree for $n_1$.}
\label{lv2-2}
\end{subfigure}
\caption{Examples with features}
\end{figure}

\begin{center}
  % \fbox{
\colorbox{teal!7}{
\begin{minipage}{0.89\linewidth}
\textbf{Level 2 Example - \texttt{bc1qah}.} 
            
\noindent\textbf{\footnotesize{GPT-4:}}
\begin{packeditemize}
\footnotesize\item{High in-degree and out-degree of transactions; Transactions like $n_1$ and $n_2$ have high in-degrees and out-degrees. \color{red}{[Average - Flawed]}}
\footnotesize\item{Significant value differences; The in- and out-value of certain transactions show significant volume and discrepancies, such as in $n_1$ and $n_2$. \color{red}{[Low]}}
\end{packeditemize}

\noindent\textbf{\footnotesize{GPT-4o:}}
\begin{packeditemize}
\footnotesize\item{High in-degree and out-degree of addresses in Layer 2; The addresses in Layer 2 have exceptionally high in-degree and out-degree values, indicating a large number of transactions. For example, address $n_{14}$ has an in-degree of 198 and out-degree of 188.  \color{red}{[High]}} 
\footnotesize\item{High in-value and out-value in transactions; The transactions in this graph involve significant amounts of value. For instance, transaction $n_1$ has an in-value of 77.29740945 and out-value of 77.29452845. \color{red}{[High]}} 
\end{packeditemize}

\end{minipage}
      }
 %  }
\end{center}

% \smallskip

% \begin{center}
%   % \fbox{
% \colorbox{teal!7}{
% \begin{minipage}{0.89\linewidth}
% \textbf{Level 2 Example - \texttt{bc1q4w}.} 
            
% \noindent\textbf{\footnotesize{GPT-4:}}
% \begin{packeditemize}
% \footnotesize\item{High out-degree in transaction nodes; A noticeable feature in this graph is the significant out-degree in some transaction nodes. For instance, $n_1$ has an out-degree of 600. \color{red}{[High]}}
% \footnotesize\item{Value discrepancy between in-value and out-value in transactions; The transaction node $n_1$ shows a discrepancy between the total incoming value (in-value: 27.69691553) and the outgoing value (out-value: 27.69184153). \color{red}{[Average - Irrelevant]}}
% \end{packeditemize}

% \noindent\textbf{\footnotesize{GPT-4o:}}
% \begin{packeditemize}
% \footnotesize\item{High out-degree transactions; The transaction node $n_1$ has an extremely high out-degree of 600, meaning it distributes funds to 600 different address nodes. \color{red}{[High]}}
% \footnotesize\item{Large volume of transactions; The transaction node $n_1$ also handles a substantial volume of in-value and out-value (27.69691553). \color{red}{[High]}} 
% \end{packeditemize}

% \end{minipage}
%       }
%  %  }
% \end{center}

\smallskip
\noindent\textit{\textbf{(i) Example - \texttt{bc1qah}.}} The responses provided by GPT-4 exhibit average or low quality. The first response correctly identifies that nodes $n_1$ and $n_2$ have high out-degrees (Fig. \ref{lv2-1}), but it flawedly describes their in-degrees. Although both nodes have higher in-degrees than many in the reference subgraph (typically in-degree is 1, while 4 and 3 for $n_1$ and $n_2$), their in-degrees are not particularly high within this graph. For example, node $n_{14}$ has an in-degree of 198, significantly exceeding that of $n_1$ and $n_2$. In the second response, most transactions have similar in-value and out-value; while the differences between in-value and out-value of $n_1$ and $n_2$ are both about 0.003, which is trivial. 

In contrast, the responses by GPT-4o are both high-quality. In the first response, $n_{14}$ does have high in-degree and out-degree (Fig.\ref{lv2-1}), while other address nodes in Layer 2, such as $n_{13}$ and $n_{19}$, also have high in-degree and out-degree. Though the second response of GPT-4o focuses on similar characteristics as GPT-4, the description of GPT-4o is accurate and reveals the high transaction values.

\smallskip
\noindent\textit{\textbf{(ii) Example - \texttt{bc1q4w}.}} The responses by GPT-4 are high-quality or average-quality. The first response is accurate and illustrates high out-degree of $n_1$ in Layer 1 (Fig.\ref{lv2-2}). Also, transaction nodes, including $n_{886}$ and $n_{1228}$ in Layer 3, have high out-degree that exceed 400. The second response is correct and demonstrates the difference between in-value and out-value. However, it is irrelevant since many transactions follow this pattern, which is ineffectual for transaction analysis.

In comparison, the responses by GPT-4o both have high quality. The first response is not only accurate and nearly identical to that of GPT-4, but also introduces the meaning of high out-degree for transaction nodes. The second response focuses on the same feature in the transaction node. It is accurate while meaningful for transaction analysis since the values are significantly high.

\vspace{0.5\baselineskip}
\begin{center}
  % \fbox{
   \colorbox{blue!7}{
         \begin{minipage}{0.89\linewidth}
            \textbf{Level 2 Findings.}
            LLMs can identify meaningful structural and transaction-value patterns in unlabeled graphs, despite occasional incorrect or irrelevant responses. GPT-4o more reliably distinguishes significant features from trivial ones, while GPT-4 more frequently misjudges their importance.

         \end{minipage}
      }
 %  }
\end{center}

\subsection{Level 3 - Contextual Interpretation}
\label{s-lv-3}

We evaluate the LLMs’ contextual interpretation capability under two settings: one using \textit{graph features} and the other using \textit{raw graphs}. In both settings, we adopt a few-shot prompting strategy, where the model is given several labeled subgraphs (or their features) as references and one unlabeled subgraph for explainable classification in each iteration of LLM query. % In future work, can consider raw graph + features

\begin{figure}[!htbp]
\centering

{\footnotesize \ref{classlegend}}
\vspace{1mm}

\begin{subfigure}[t]{0.48\linewidth}
\centering
\resizebox{\linewidth}{4.5cm}{%
\begin{tikzpicture}
    \begin{axis}[
        ybar,
        width=\linewidth,
        height=8cm,
        ymin=0, ymax=80,
        grid=major,
        grid style=dashed,
        bar width=6pt,
        enlarge x limits=0.2,
        xtick=data,
        symbolic x coords={GPT-3.5, GPT-4, GPT-4o, DeepSeek},
        xticklabel style={rotate=0, anchor=north},
        x tick label style={align=center},
        x=2.0cm,
        legend to name=classlegend,
        legend columns=5,
        legend style={font=\footnotesize}
    ]
        \addplot[fill=blue!30] coordinates {
            (GPT-3.5,39.83) (GPT-4,46.07) (GPT-4o,42.85) (DeepSeek,46.68)
        };
        \addplot[fill=red!30] coordinates {
            (GPT-3.5,70.02) (GPT-4,67.20) (GPT-4o,71.02) (DeepSeek,72.43)
        };
        \addplot[fill=green!30] coordinates {
            (GPT-3.5,31.53) (GPT-4,37.42) (GPT-4o,39.84) (DeepSeek,33.22)
        };
        \addplot[fill=brown!30] coordinates {
            (GPT-3.5,35.83) (GPT-4,44.60) (GPT-4o,47.45) (DeepSeek,44.65)
        };
        \addplot[fill=purple!30] coordinates {
            (GPT-3.5,31.96) (GPT-4,35.40) (GPT-4o,35.18) (DeepSeek,36.56)
        };
        \legend{Accuracy, Top-3 Acc., Precision, Recall, F1}
    \end{axis}
\end{tikzpicture}%
}
\caption{Feature-based}
\label{lv3-01}
\end{subfigure}%
\hfill%
\begin{subfigure}[t]{0.48\linewidth}
\centering
\resizebox{\linewidth}{4.5cm}{%
\begin{tikzpicture}
    \begin{axis}[
        ybar,
        width=\linewidth,
        height=8cm,
        ymin=0, ymax=80,
        grid=major,
        grid style=dashed,
        bar width=6pt,
        enlarge x limits=0.2,
        xtick=data,
        symbolic x coords={GPT-4, GPT-4o, DeepSeek},
        xticklabel style={rotate=0, anchor=north},
        x tick label style={align=center},
        x=2.5cm
    ]
        \addplot[fill=blue!30] coordinates {
            (GPT-4,18.81) (GPT-4o,50.49) (DeepSeek,41.58)
        };
        \addplot[fill=red!30] coordinates {
            (GPT-4,48.51) (GPT-4o,68.31) (DeepSeek,64.35)
        };
        \addplot[fill=green!30] coordinates {
            (GPT-4,27.61) (GPT-4o,63.33) (DeepSeek,48.12)
        };
        \addplot[fill=brown!30] coordinates {
            (GPT-4,21.40) (GPT-4o,53.95) (DeepSeek,41.29)
        };
        \addplot[fill=purple!30] coordinates {
            (GPT-4,16.37) (GPT-4o,50.14) (DeepSeek,35.54)
        };
    \end{axis}
\end{tikzpicture}%
}
\caption{Raw graph-based}
\label{lv3-02}
\end{subfigure}

\caption{Classification via different LLMs. The x-axis indicates LLM models and the y-axis indicates percentage (\%). The five bars represent accuracy, top-3 accuracy, precision, recall, and F1 score, respectively.}
\label{fig:classification_side_by_side}
\end{figure}
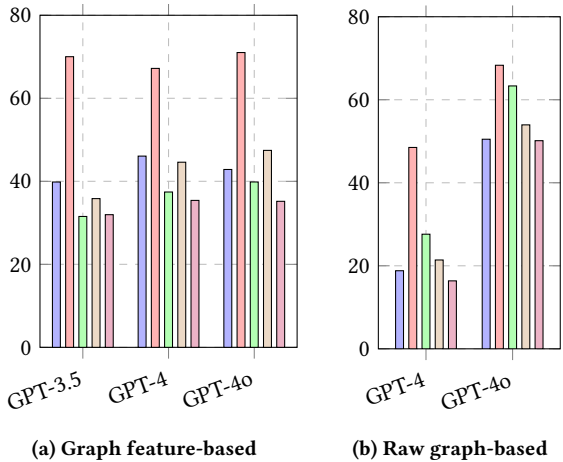

\smallskip
\noindent\textbf{Graph Feature-based Classification.} In this setting, we provide LLMs with subgraph features and their textual descriptions, including both labeled and unlabeled samples. Specifically, we use the ten most important features identified in BABD \cite{10375557}. We then evaluate the models’ performance across classes using accuracy, macro precision, recall, and F1 scores.

% The metrics (overall)
The overall accuracy of LLMs (Fig.\ref{lv3-01}) ranges from 39.83\% to 46.68\%, with similarly limited performance in macro precision, recall, and F1 scores. Although these results appear modest, they are partly attributable to the scarcity of reference samples, which constrains precise classification. Notably, the top-3 accuracy remains high, i.e., 67.20\% to 72.43\%, indicating that while exact class prediction is challenging under limited data, LLMs can still reliably narrow down the correct category within the top three candidates.

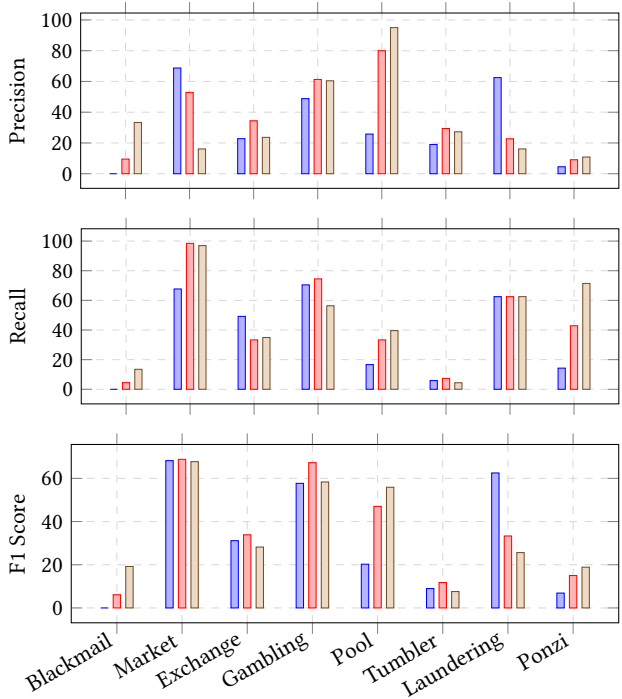
\begin{figure}[t]
    \centering
    % Precision subfigure
    \begin{subfigure}[t]{0.47\textwidth}
        \centering
        \resizebox{\linewidth}{!}{
        \begin{tikzpicture}
            \begin{axis}[
                ybar,
                bar width=0.1cm,
                width=9cm, % Plot width
                height=4cm, % Plot height
                symbolic x coords={Blackmail, Market, Exchange, Gambling, Pool, Tumbler, Laundering, Ponzi},
                ytick={0,20,40,60,80,100},
                xtick=data,
                xticklabels={}, 
                grid=major,
                grid style={dashed, gray!30}, 
                legend style={at={(0.5,1.4)}, anchor=north, legend columns=-1},
                enlarge x limits=0.1,
                ylabel={Precision},
            ]
                % GPT-3.5, GPT-4, GPT-4o
                \addplot coordinates {(Blackmail,0) (Market,68.75) (Exchange,22.79) (Gambling,48.83) (Pool,25.80) (Tumbler,19.04) (Laundering,62.50) (Ponzi,4.54)};
                \addplot coordinates {(Blackmail,9.52) (Market,52.89) (Exchange,34.42) (Gambling,61.32) (Pool,80.00) (Tumbler,29.41) (Laundering,22.72) (Ponzi,9.09)};
                \addplot coordinates {(Blackmail,33.33) (Market,16.12) (Exchange,23.65) (Gambling,60.43) (Pool,95.00) (Tumbler,27.27) (Laundering,16.12) (Ponzi,10.86)};
                % DeepSeek (feature)
                \addplot[ybar,draw=black,fill=black,fill opacity=0.3] coordinates {(Blackmail,9.67) (Market,50.39) (Exchange,34.92) (Gambling,55.32) (Pool,69.23) (Tumbler,0.00) (Laundering,35.71) (Ponzi,10.52)};
                \legend{GPT-3.5, GPT-4, GPT-4o, DeepSeek}
            \end{axis}
        \end{tikzpicture}
        }
        \vspace{-0.3in}
    \end{subfigure}
\hfill
    % Recall subfigure
    \begin{subfigure}[t]{0.47\textwidth}
        \centering
        \resizebox{\linewidth}{!}{
        \begin{tikzpicture}
            \begin{axis}[
                ybar,
                bar width=0.1cm,
                width=9cm,
                height=4cm,
                symbolic x coords={Blackmail, Market, Exchange, Gambling, Pool, Tumbler, Laundering, Ponzi},
                ytick={0,20,40,60,80,100},
                xtick=data,
                xticklabels={}, 
                grid=major,
                grid style={dashed, gray!30}, 
                enlarge x limits=0.1,
                ylabel={Recall}
            ]
                % GPT-3.5, GPT-4, GPT-4o
                \addplot coordinates {(Blackmail,0) (Market,67.69) (Exchange,49.20) (Gambling,70.46) (Pool,16.66) (Tumbler,5.88) (Laundering,62.50) (Ponzi,14.28)};
                \addplot coordinates {(Blackmail,4.49) (Market,98.46) (Exchange,33.33) (Gambling,74.49) (Pool,33.33) (Tumbler,7.35) (Laundering,62.50) (Ponzi,42.85)};
                \addplot coordinates {(Blackmail,13.48) (Market,96.92) (Exchange,34.92) (Gambling,56.37) (Pool,39.58) (Tumbler,4.41) (Laundering,62.50) (Ponzi,71.42)};
                % DeepSeek (feature)
                \addplot[ybar,draw=black,fill=black,fill opacity=0.3] coordinates {(Blackmail,3.37) (Market,98.46) (Exchange,34.92) (Gambling,73.15) (Pool,56.25) (Tumbler,0.00) (Laundering,62.50) (Ponzi,28.57)};
            \end{axis}
        \end{tikzpicture}
        }
        \vspace{-0.3in}
    \end{subfigure}
\hfill
    % F1 Score subfigure
    \begin{subfigure}[t]{0.47\textwidth}
        \centering
        \resizebox{\linewidth}{!}{
        \begin{tikzpicture}
            \begin{axis}[
                ybar,
                bar width=0.1cm,
                width=9cm,
                height=4cm,
                symbolic x coords={Blackmail, Market, Exchange, Gambling, Pool, Tumbler, Laundering, Ponzi},
                ytick={0,20,40,60,80,100},
                xtick=data,
                xticklabel style={rotate=30, anchor=east},
                grid=major,
                grid style={dashed, gray!30}, 
                enlarge x limits=0.1,
                ylabel={F1 Score},
            ]
                % GPT-3.5, GPT-4, GPT-4o
                \addplot coordinates {(Blackmail,0) (Market,68.21) (Exchange,31.15) (Gambling,57.69) (Pool,20.25) (Tumbler,8.98) (Laundering,62.50) (Ponzi,6.89)};
                \addplot coordinates {(Blackmail,6.10) (Market,68.81) (Exchange,33.87) (Gambling,67.27) (Pool,47.05) (Tumbler,11.76) (Laundering,33.33) (Ponzi,15.00)};
                \addplot coordinates {(Blackmail,19.20) (Market,67.74) (Exchange,28.20) (Gambling,58.33) (Pool,55.88) (Tumbler,7.59) (Laundering,25.64) (Ponzi,18.86)};
                % DeepSeek (feature)
                \addplot[ybar,draw=black,fill=black,fill opacity=0.3] coordinates {(Blackmail,5.00) (Market,66.66) (Exchange,34.92) (Gambling,63.00) (Pool,62.06) (Tumbler,0.00) (Laundering,45.45) (Ponzi,15.38)};
            \end{axis}
        \end{tikzpicture}
        }
        \vspace{-0.43in}
    \end{subfigure}

    \caption{LLMs' performance in contextual interpretation using \textcolor{teal}{graph features} (x axis for category, y for rate (\%); GPT-3.5 in blue bar, GPT-4 in red, GPT-4o in brown, DeepSeek in gray).}
    \label{fig:capability-graphfeature}
\end{figure}

% The metrics (class)
Beyond the overall metrics, Fig.~\ref{fig:capability-graphfeature} reveals uneven, category-dependent strengths. GPT-4 and GPT-4o generally set the pace, pairing standout precision in mining pools at 80.00\% and 95.00\% with near-saturated recall on the darknet market at 98.46\% and 96.92\%. DeepSeek remains competitive where a recall-weighted balance matters, yielding the strongest F1 on pools and a slight edge on exchange. GPT-3.5, though weaker overall, leads on money laundering because all models share the same recall there and its precision is highest, producing the top F1. This suggests that LLM performance is complementary rather than absolute, with relative strengths distributed unevenly across categories in this context.

% Compared among LLMs
When comparing models, GPT-4o attains the highest recall of 47.45\% and macro precision of 39.84\%, suggesting greater consistency across categories. DeepSeek records the strongest overall accuracy at 46.68\%, top-3 accuracy at 72.43\%, and F1 score of 36.56\%, though accompanied by notably lower precision of 33.22\%. GPT-4 yields comparatively balanced precision and recall, whereas GPT-3.5 trails in most metrics. Taken together, these results show that performance advantages are distributed unevenly, with different models favoring different trade-offs.

Focusing specifically on the GPT series, GPT-4o demonstrates better stability, with fewer categories having precision, recall, or F1 scores significantly low. These results suggest that despite updates leading to marginal improvements in graph feature-based classification, GPT-4o still offers mild but limited advantages over GPT-3.5 and GPT-4.

\smallskip
\noindent\textbf{Raw Graph-based Classification.} For raw graph-based classification, we rely on fewer subgraphs, as the higher computational cost limits the scale compared to feature-based setting.

\begin{comment}
\begin{table}[!htbp]
    \centering
    \caption{Graph Feature-based Classification Results}
    \footnotesize
    \vspace{-1\baselineskip}
%     \setlength{\tabcolsep}{1.1em}
% \renewcommand{\arraystretch}{1.2}
    \begin{tabular}{cccccc}
        \toprule
        \textbf{Model} & \textbf{Accuracy} & \textbf{Top-3 Accuracy} & \textbf{Precision} & \textbf{Recall} & \textbf{F1 Score} \\ 
        \midrule
        \textbf{GPT-3.5} & 39.83\% & 70.02\% & 31.53\% & 35.83\% & 31.96\% \\ 
        \textbf{GPT-4}   & 46.07\% & 67.20\% & 37.42\% & 44.60\% & 35.40\% \\ 
        \textbf{GPT-4o}  & 42.85\% & 71.02\% & 39.84\% & 47.45\% & 35.18\% \\ 
        \bottomrule
    \end{tabular}
    \label{lv3-01}
    \begin{tablenotes}
    \scriptsize
    \item \hspace{0em} *Precision, recall, and F1 score are macro-averages (same in Table \ref{lv3-02})
    \end{tablenotes}
    \end{table}

\begin{table}[!htbp]
    \centering
    \caption{Raw Graph-based Classification Results}
    \footnotesize
    \vspace{-1\baselineskip}
% \setlength{\tabcolsep}{1.1em}
% \renewcommand{\arraystretch}{1.2}
    \begin{tabular}{cccccc}
        \toprule
        \textbf{Model}  & \textbf{Accuracy} & \textbf{Top-3 Accuracy} & \textbf{Precision} & \textbf{Recall} & \textbf{F1 Score} \\ 
        \midrule
        \textbf{GPT-4}  & 18.81\% & 48.51\% & 27.61\% & 21.40\% & 16.37\% \\ 
        \textbf{GPT-4o} & 50.49\% & 68.31\% & 63.33\% & 53.95\% & 50.14\% \\ 
        \bottomrule
    \end{tabular}
    \label{lv3-02}
\end{table}
\end{comment}

% The metrics (overall)
The overall accuracy of GPT-4 (Fig.\ref{lv3-02}) is significantly lower than desired. In contrast, GPT-4o surpasses all feature-based classification tasks, achieving an accuracy of 50.49\%. The macro precision, recall, and F1 scores of GPT-4 on raw graph-based classification tasks fall below those of GPT-3.5, GPT-4, and GPT-4o on graph feature-based classification tasks. Conversely, the metrics for GPT-4o on graph-based classification tasks are notably higher than those on graph feature-based classification tasks, all exceeding 50.14\%. As for top-3 accuracy, GPT-4 is still lower than average, while GPT-4o achieves similar performance on raw graph-based classification compared with those on feature-based. While DeepSeek achieves competitive accuracy of 41.58\% and F1 score of 35.54\% using raw graphs, clearly outperforming GPT-4 yet still trailing GPT-4o.

\definecolor{mydeepbrown}{RGB}{90,60,30}

\begin{figure}[t]
    \centering
    % Precision subfigure
    \begin{subfigure}[t]{0.47\textwidth}
        \centering
        \resizebox{0.99\linewidth}{!}{
        \begin{tikzpicture}
            \begin{axis}[
                ybar,
                bar width=0.12cm,
                width=8.5cm, % Plot width
                height=4cm, % Plot height
                symbolic x coords={Blackmail, Market, Exchange, Gambling, Pool, Tumbler, Laundering, Ponzi},
                ytick={0,20,40,60,80,100},
                xtick=data,
                xticklabels={}, % 设置为空 顺序要在x coords后面
                grid=major,
                grid style={dashed, gray!30}, 
                legend style={at={(0.5,1.4)}, anchor=north, legend columns=-1},
                enlarge x limits=0.15,
                ylabel={Precision}
            ]
                % GPT-4, GPT-4o
                \addplot[ybar,draw=red,fill=red,fill opacity=0.3] coordinates {(Blackmail,0) (Market,100) (Exchange,16.66) (Gambling,0) (Pool,50) (Tumbler,14.28) (Laundering,40) (Ponzi,0)};
                \addplot[ybar,draw=mydeepbrown,fill=brown,fill opacity=0.3] coordinates {(Blackmail,0) (Market,100) (Exchange,26.66) (Gambling,100) (Pool,43.75) (Tumbler,56.25) (Laundering,80) (Ponzi,100)};
                % DeepSeek (raw) 
                \addplot[ybar,draw=black,fill=black,fill opacity=0.3] coordinates {(Blackmail,100.0) (Market,93.75) (Exchange,47.05) (Gambling,72.72) (Pool,16.66) (Tumbler,40.00) (Laundering,14.81) (Ponzi,0.00)};
                \legend{GPT-4, GPT-4o, DeepSeek}
            \end{axis}
        \end{tikzpicture}
        }
        \vspace{-0.13in}
     %   \caption{Precision}
    \end{subfigure}
\hfill
    % Recall subfigure
    \begin{subfigure}[t]{0.47\textwidth}
        \centering
        \resizebox{0.99\linewidth}{!}{
        \begin{tikzpicture}
            \begin{axis}[
                ybar,
                bar width=0.12cm,
                width=8.5cm, % Plot width
                height=4cm, % Plot height
                symbolic x coords={Blackmail, Market, Exchange, Gambling, Pool, Tumbler, Laundering, Ponzi},
                ytick={0,20,40,60,80,100},
                xtick=data,
                xticklabels={}, % 设置为空 顺序要在x coords后面
                grid=major,
                grid style={dashed, gray!30}, 
                enlarge x limits=0.15,
                ylabel={Recall}
            ]
                % GPT-4, GPT-4o
                \addplot[ybar,draw=red,fill=red,fill opacity=0.3] coordinates {(Blackmail,0) (Market,6.25) (Exchange,21.42) (Gambling,0) (Pool,33.33) (Tumbler,76.92) (Laundering,33.33) (Ponzi,0)};
                \addplot[ybar,draw=mydeepbrown,fill=brown,fill opacity=0.3] coordinates {(Blackmail,0) (Market,81.25) (Exchange,85.71) (Gambling,17.64) (Pool,77.77) (Tumbler,69.23) (Laundering,66.66) (Ponzi,33.33)};
                % DeepSeek (raw) —— 统一为黑色
                \addplot[ybar,draw=black,fill=black,fill opacity=0.3] coordinates {(Blackmail,5.88) (Market,93.75) (Exchange,57.14) (Gambling,47.05) (Pool,44.44) (Tumbler,15.38) (Laundering,66.66) (Ponzi,0.00)};
            \end{axis}
        \end{tikzpicture}
        }
        \vspace{-0.13in}
    %    \caption{Recall}
    \end{subfigure}
\hfill
    % F1 Score subfigure
    \begin{subfigure}[t]{0.47\textwidth}
        \centering
         \resizebox{0.99\linewidth}{!}{
        \begin{tikzpicture}
            \begin{axis}[
                ybar,
                bar width=0.12cm,
                width=8.5cm, % Plot width
                height=4cm, % Plot height
                xtick=data,
                xticklabel style={rotate=30, anchor=east},
                symbolic x coords={Blackmail, Market, Exchange, Gambling, Pool, Tumbler, Laundering, Ponzi},
                ytick={0,20,40,60,80,100},
                grid=major,
                grid style={dashed, gray!30}, 
                enlarge x limits=0.15,
                ylabel={F1 Score}
            ]
                % GPT-4, GPT-4o
                \addplot[ybar,draw=red,fill=red,fill opacity=0.3] coordinates {(Blackmail,0) (Market,11.76) (Exchange,18.75) (Gambling,0) (Pool,40) (Tumbler,24.09) (Laundering,36.36) (Ponzi,0)};
                \addplot[ybar,draw=mydeepbrown,fill=brown,fill opacity=0.3] coordinates {(Blackmail,0) (Market,89.65) (Exchange,40.67) (Gambling,30) (Pool,56) (Tumbler,62.06) (Laundering,72.72) (Ponzi,50)};
                % DeepSeek (raw) 
                \addplot[ybar,draw=black,fill=black,fill opacity=0.3] coordinates {(Blackmail,11.11) (Market,93.75) (Exchange,51.61) (Gambling,57.14) (Pool,24.24) (Tumbler,22.22) (Laundering,24.24) (Ponzi,0.00)};
            \end{axis}
        \end{tikzpicture}
        }
       \vspace{-0.3in}
     %   \caption{F1 Score}
    \end{subfigure}
    \caption{LLMs' performance in contextual interpretation using \textcolor{teal}{raw graphs}  (x axis for category, y for rate (\%); GPT-4 in red, GPT-4o in brown, DeepSeek in gray).}
    \label{fig:capability-rawdata}
\end{figure}
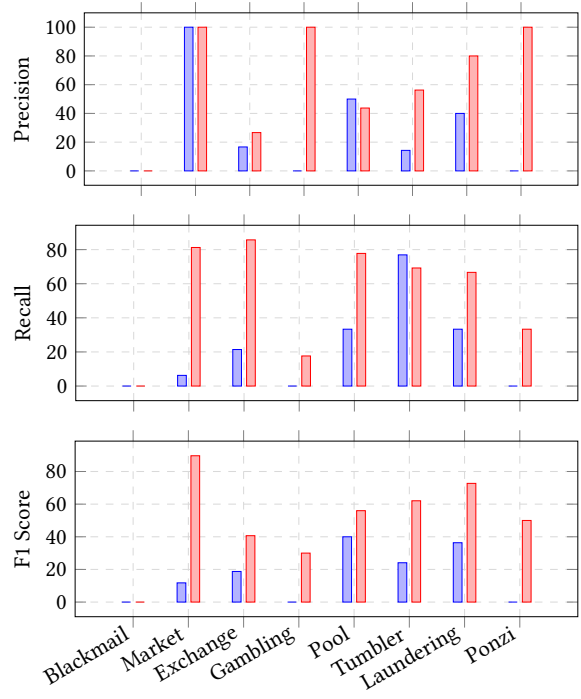

For the specific class as shown in Figure~\ref{fig:capability-rawdata}, GPT-4o performs better than GPT-4 in almost all metrics, in some classes such as darknet market, the differences of F1 scores are huge. One remarkable exception is that both GPT-4 and GPT-4o recorded zero scores in all metrics on blackmail; however, for GPT-4, classes including gambling and Ponzi scheme recorded zero scores across all metrics, reflecting a huge failure in identifying relevant instances. In contrast, DeepSeek achieves competitive performance in several categories, particularly darknet market, exchange, and gambling, where its F1 scores are markedly higher than GPT-4, although it still struggles in classes such as pool and Ponzi scheme, where GPT-4o maintains clear advantages.

To sum up, GPT-4o consistently achieves substantially higher performance than GPT-4 across all metrics in raw graph-based tasks and outperforms it in most specific classes. DeepSeek also delivers competitive results, showing notable strengths in categories such as darknet market and gambling, though it remains less stable than GPT-4o overall. These findings suggest that model optimization may enhance raw graph-based classification, while different LLM architectures may offer complementary advantages across categories.

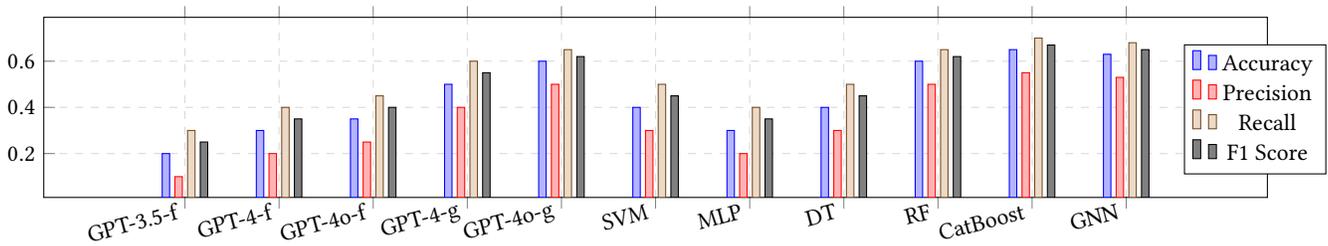
\begin{figure}
\centering
\resizebox{0.5\textwidth}{!}{
\begin{tikzpicture}
    \begin{axis}[
        ybar,
        bar width=0.1cm,
        width=\linewidth,
        height=4cm,
        x=1.1cm,               
        enlarge x limits=0.05, 
        grid=major,
        grid style={dashed, gray!30}, 
        legend style={at={(0.5,1.1)}, anchor=south, legend columns=4},
        symbolic x coords={GPT-3.5-f, GPT-4-f, GPT-4o-f, GPT-4-g, GPT-4o-g, DeepSeek-f, DeepSeek-g, SVM, MLP, DT, RF, CatBoost, GNN},
        xtick=data,
        x tick label style={rotate=30,anchor=east,font=\footnotesize},
        nodes near coords align={vertical},
        %y轴显示
        ymin=0, ymax=0.8,
        ytick={0,0.2,0.4,0.6},
        yticklabels={0,20,40,60},
    ]
    % Accuracy
    \addplot coordinates {
        (GPT-3.5-f, 0.3983) (GPT-4-f, 0.4607) (GPT-4o-f, 0.4285) 
        (GPT-4-g, 0.1881) (GPT-4o-g, 0.5049) 
        (DeepSeek-f, 0.4668) (DeepSeek-g, 0.4158)
        (SVM, 0.1120) (MLP, 0.1800) (DT, 0.5740) (RF, 0.6000) (CatBoost, 0.5880) (GNN, 0.6840)};
    % Precision
    \addplot coordinates {
        (GPT-3.5-f, 0.3153) (GPT-4-f, 0.3742) (GPT-4o-f, 0.3984) 
        (GPT-4-g, 0.2761) (GPT-4o-g, 0.6333) 
        (DeepSeek-f, 0.3322) (DeepSeek-g, 0.4812)
        (SVM, 0.1989) (MLP, 0.0684) (DT, 0.5075) (RF, 0.5374) (CatBoost, 0.5311) (GNN, 0.6110)};
    % Recall
    \addplot coordinates {
        (GPT-3.5-f, 0.3583) (GPT-4-f, 0.4460) (GPT-4o-f, 0.4745) 
        (GPT-4-g, 0.2140) (GPT-4o-g, 0.5395) 
        (DeepSeek-f, 0.4465) (DeepSeek-g, 0.4129)
        (SVM, 0.2246) (MLP, 0.2064) (DT, 0.5845) (RF, 0.6775) (CatBoost, 0.6783) (GNN, 0.5560)};
    % F1
    \addplot coordinates {
        (GPT-3.5-f, 0.3196) (GPT-4-f, 0.3540) (GPT-4o-f, 0.3518) 
        (GPT-4-g, 0.1637) (GPT-4o-g, 0.5014) 
        (DeepSeek-f, 0.3656) (DeepSeek-g, 0.3554)
        (SVM, 0.1242) (MLP, 0.1013) (DT, 0.5056) (RF, 0.5421) (CatBoost, 0.5337) (GNN, 0.5820)};
    \legend{Accuracy, Precision, Recall, F1 Score}
    \end{axis}
\end{tikzpicture}
}
\caption{Evaluation on different models (x axis for category, y for corresponding rates (\%))}
\label{fig:models_comparison}
\end{figure}

\smallskip
\noindent\textbf{Comparison with Traditional Models.} As illustrated in Fig.\ref{fig:models_comparison}, in the context of a very small number of reference samples, LLMs based on graph features generally perform significantly better than support vector machine (SVM) and MLP, yet remain inferior compared to DT, RF, CatBoost, and GNN. Among feature-based LLMs, DeepSeek delivers the strongest results, with accuracy 46.68\% and F1 score 36.56\% surpassing GPT-3.5 and GPT-4, though still behind tree models and GNN.

For raw-graph inputs, performance varies widely. GPT-4 only slightly outperforms support vector machine and multilayer perceptron, whereas GPT-4o achieves substantial improvements, with accuracy 50.49\% and precision 63.33\% comparable to tree models. This high precision is crucial for illegal address detection, as misclassifying legitimate addresses as illegal could lead to severe consequences such as account suspension or fund freezing. DeepSeek, however, lags behind GPT-4o in both accuracy and F1, though it still demonstrates relatively strong precision 48.12\%, close to some tree models. 

Overall, GPT-4o analyzing sampled raw graphs attains performance close to carefully engineered tree-based methods, while DeepSeek shows that feature-driven methods can also be highly competitive.

\smallskip
\noindent\textbf{Analysis of Classification Explanation.} Unlike traditional results of classification tasks, LLM-based outcomes can include detailed explanations that may be valuable for further analysis. However, our initial investigation indicates that these explanations are not always accurate. We choose two near-accurate explanations for  \texttt{1Bsjsa\footnote{\texttt{1BsjsaHST2Qohs8ZHxNHeZ1UfWhtxoKHEN}}} and \texttt{124mpe\footnote{\texttt{124mpePGM2vEqcHUK96wQXur9vS7Vn7Kdj}}} from GPT-4o to show the potential of LLMs and how they utilize different types of data.

% \smallskip

\begin{center}
  % \fbox{
\colorbox{teal!17}{
\begin{minipage}{0.89\linewidth}
\textbf{Level 3 Example - \texttt{1Bsjsa}.} 

\noindent\footnotesize{\textbf{Ground Truth:} Ponzi}\\
\noindent\footnotesize{\textbf{Feature-based Result:} Ponzi, blackmail, tumbler}\\
\noindent\textbf{\footnotesize{Reason:}}

\begin{packeditemize}
\footnotesize\item{\emph{CI3a32-2} value is extremely high, which is a distinctive characteristic of Ponzi graphs.}
\footnotesize\item{\emph{S7} is also higher than average, another indicative feature for Ponzi. Additionally, \emph{S5} and \emph{S6} align well with Ponzi characteristics.}
\footnotesize\item{Blackmail and tumbler categories are also considered due to similar \emph{S2-2} and \emph{S6} values.}
\end{packeditemize}

\noindent\footnotesize{\textbf{Graph-based Result:} Ponzi, tumbler, pool}\\
\noindent\textbf{\footnotesize{Reason:}}

\begin{packeditemize}
\footnotesize\item{The graph has a structure where a single address node ($n_0$) has a high degree of both incoming and outgoing transactions (300 in-degree and 197 out-degree). This pattern is indicative of schemes like Ponzi, tumbler, and pool, where a central node is involved in numerous transactions with other nodes.}
\footnotesize\item{The high number of small-value transactions and the presence of multiple transaction nodes with varying degrees further support these categories.} 
\end{packeditemize}

\end{minipage}
      }
 %  }
\end{center}

% \smallskip

% \begin{center}
%   % \fbox{
% \colorbox{teal!17}{
% \begin{minipage}{0.89\linewidth}
% \textbf{Level 3 Example - \texttt{124mpe}.} 

% \noindent\footnotesize{\textbf{Ground Truth:} gambling}\\
% \noindent\footnotesize{\textbf{Feature-based Result:} gambling, darknet market, exchange}\\
% \noindent\textbf{\footnotesize{Reason:}}

% \begin{packeditemize}
% \footnotesize\item{The high value of \emph{PTIa41-2} is a strong indicator of gambling as similar high values are observed in reference graphs.}
% \footnotesize\item{The \emph{S2-2}, \emph{S1-6}, \emph{S1-2}, and \emph{S3} values are within ranges observed in darknet market and exchange categories, making them the next most probable categories.}
% \end{packeditemize}

% \noindent\footnotesize{\textbf{Graph-based Result:} gambling, pool, exchange}\\
% \noindent\textbf{\footnotesize{Reason:}}

% \begin{packeditemize}
% \footnotesize\item{The structure and value distributions of the graph suggest that it is most likely a gambling, pool, or exchange category. The presence of large in-degree transactions and multiple out-degree addresses is characteristic of gambling and pool activities, as seen in the reference graphs.}
% \footnotesize\item{The significant transaction values and multiple address connections also align with the patterns found in exchange-related graphs} 
% \end{packeditemize}

% \end{minipage}
%       }
%  %  }
% \end{center}

As demonstrated above, GPT-4o can almost accurately categorize \texttt{1Bsjsa} from two perspectives. From a graph feature perspective, \emph{CI3a32-2} and \emph{S7} are the primary factors influencing GPT-4o to classify it as a Ponzi address. \emph{S5} and \emph{S6} also play significant roles, where the values of these features compared with the reference graph features are crucial in this determination. From a raw graph perspective, the high in-degree and out-degree with massive small-value transactions is the key reason for the conclusion. 

Similarly, GPT-4o is capable of accurately identifying \texttt{124mpe} from two perspectives. From a graph feature perspective, \emph{PTIa41-2} is the primary factor influencing GPT-4o to classify it as a gambling address. \emph{S2-2}, \emph{S1-6}, \emph{S1-2}, and \emph{S3} also play significant roles. The values of these features compared with reference graph features are crucial in determining its classification. From a raw graph perspective, the high in-degree transactions and multiple out-degree addresses with significant transaction values are key reasons for this conclusion, strongly indicating gambling, pool, or exchange activities.

These examples illustrate that for feature-based classification, adequate quantity and quality samples with labels are required to improve classification consequences since the selected features are already processed. In contrast, for graph-based classification, the quality of samples themselves seems to be also essential. The unsuitable compression of the graphs leading to much information loss may negatively affect the results. Also, the chain of thought (CoT) prompting \cite{wei2022chain} could be used for further explanations, with detailed reasoning and better classification interpretability.

\vspace{0.5\baselineskip}
\begin{center}
  % \fbox{
   \colorbox{blue!7}{
         \begin{minipage}{0.89\linewidth}
            \textbf{Level 3 Findings.}
            LLMs achieve relatively high top-3 classification accuracy with limited labeled data and provide plausible explanations. However, moderate top-1 accuracy and weaker performance than specialized tree models and GNNs limit their effectiveness in precise classification, making them more suitable for exploratory analysis.
         \end{minipage}
      }
 %  }
\end{center}

%==============================================
\section{Discussion}
\label{sec-discu}
%==============================================

\subsection{Graph Representation and Token Efficiency}

\begin{table*}[t]
\centering
\caption{LLM capability on foundational metrics across graph formats and models}
\label{lv1-extra}
\footnotesize
\renewcommand{\arraystretch}{0.90}
\setlength{\tabcolsep}{4pt}
\resizebox{\linewidth}{!}{
\begin{tabular}{lcccccccccccc}
\toprule
 & \multicolumn{4}{c}{\textbf{GPT-4o}} & \multicolumn{4}{c}{\textbf{LLaMA}} & \multicolumn{4}{c}{\textbf{DeepSeek}} \\
\cmidrule(lr){2-5}\cmidrule(lr){6-9}\cmidrule(lr){10-13}
\textbf{Metrics} & \textbf{LLM4TG} & \textbf{GEXF} & \textbf{GML} & \textbf{GraphML}
                & \textbf{LLM4TG} & \textbf{GEXF} & \textbf{GML} & \textbf{GraphML}
                & \textbf{LLM4TG} & \textbf{GEXF} & \textbf{GML} & \textbf{GraphML} \\
\midrule
\emph{struct\_correctness}     & 100.00\% & 95.83\% & 95.83\% & 100.00\% & 100.00\% & 87.50\% & 95.83\% & 87.50\% & 100.00\% & 95.83\% & 91.67\% & 95.83\% \\
\midrule
\emph{global\_in\_degree}      & 41.67\%  & 78.26\% & 78.26\% & 75.00\%  & 54.17\%  & 71.43\% & 69.57\% & 76.19\% & 50.00\%  & 69.57\% & 50.00\% & 60.87\% \\
\emph{global\_out\_degree}     & 62.50\%  & 60.87\% & 56.52\% & 54.17\%  & 50.00\%  & 71.43\% & 73.91\% & 76.19\% & 58.33\%  & 47.83\% & 36.36\% & 39.13\% \\
\emph{global\_in\_value}       & 54.17\%  & 30.43\% & 52.17\% & 25.00\%  & 33.33\%  & 33.33\% & 21.74\% & 47.62\% & 45.83\%  & 56.52\% & 45.45\% & 47.83\% \\
\emph{global\_out\_value}      & 25.00\%  & 26.09\% & 30.43\% & 25.00\%  & 45.83\%  & 9.52\%  & 8.70\%  & 19.05\% & 37.50\%  & 26.09\% & 22.73\% & 26.09\% \\
\emph{global\_diff\_degree}    & 37.50\%  & 43.48\% & 43.48\% & 41.67\%  & 33.33\%  & 47.62\% & 43.48\% & 52.38\% & 29.17\%  & 30.43\% & 18.18\% & 34.78\% \\
\emph{global\_diff\_value}     & 8.33\%   & 8.70\%  & 8.70\%  & 8.33\%   & 25.00\%  & 0.00\%  & 4.35\%  & 4.76\%  & 16.67\%  & 8.70\%  & 4.55\%  & 4.35\%  \\
\midrule
\emph{node\_in\_degree}        & 73.75\%  & 75.22\% & 75.65\% & 73.33\%  & 73.33\%  & 66.19\% & 60.87\% & 69.05\% & 75.42\%  & 73.91\% & 67.27\% & 68.26\% \\
\emph{node\_out\_degree}       & 75.83\%  & 82.17\% & 80.43\% & 79.17\%  & 76.25\%  & 67.14\% & 54.78\% & 70.48\% & 77.50\%  & 76.96\% & 72.27\% & 70.87\% \\
\emph{node\_in\_value}         & 75.42\%  & 70.43\% & 71.74\% & 67.92\%  & 73.75\%  & 69.05\% & 58.26\% & 68.10\% & 77.50\%  & 69.13\% & 67.27\% & 67.83\% \\
\emph{node\_out\_value}        & 77.08\%  & 74.78\% & 71.74\% & 71.67\%  & 78.33\%  & 55.71\% & 55.65\% & 58.57\% & 79.17\%  & 70.00\% & 68.64\% & 65.65\% \\
\emph{node\_special\_info\_a}  & 100.00\% & 23.08\% & 14.66\% & 11.67\%  & 99.17\%  & 47.17\% & 47.86\% & 44.34\% & 100.00\% & 45.69\% & 46.43\% & 49.57\% \\
\emph{node\_special\_info\_t}  & 71.67\%  & 81.42\% & 64.91\% & 79.17\%  & 78.33\%  & 76.92\% & 72.57\% & 72.12\% & 83.90\%  & 63.16\% & 64.81\% & 68.14\% \\
\bottomrule
\end{tabular}
}
\end{table*}

\noindent\textbf{Cross-Format Evaluation.}
Table~\ref{lv1-extra} compares LLM4TG with GEXF, GML, and GraphML across three LLMs. LLM4TG achieves 100\% structural correctness and consistently strong performance in extracting node attributes. In particular, its accuracy on \emph{node\_special\_info\_a} exceeds 99\% across all models, substantially outperforming conventional formats. However, the advantages are less consistent for other metrics. Conventional formats occasionally outperform LLM4TG on node-level and global metrics, particularly \emph{global\_in\_degree}. Moreover, \emph{global\_diff\_value} remains challenging across all formats, suggesting limitations in cross-node numerical reasoning. These results indicate that LLM4TG improves the accessibility of graph attributes without consistently improving all graph reasoning tasks.

\begin{figure}[!]
    \centering
    \includegraphics[width=0.8\linewidth]{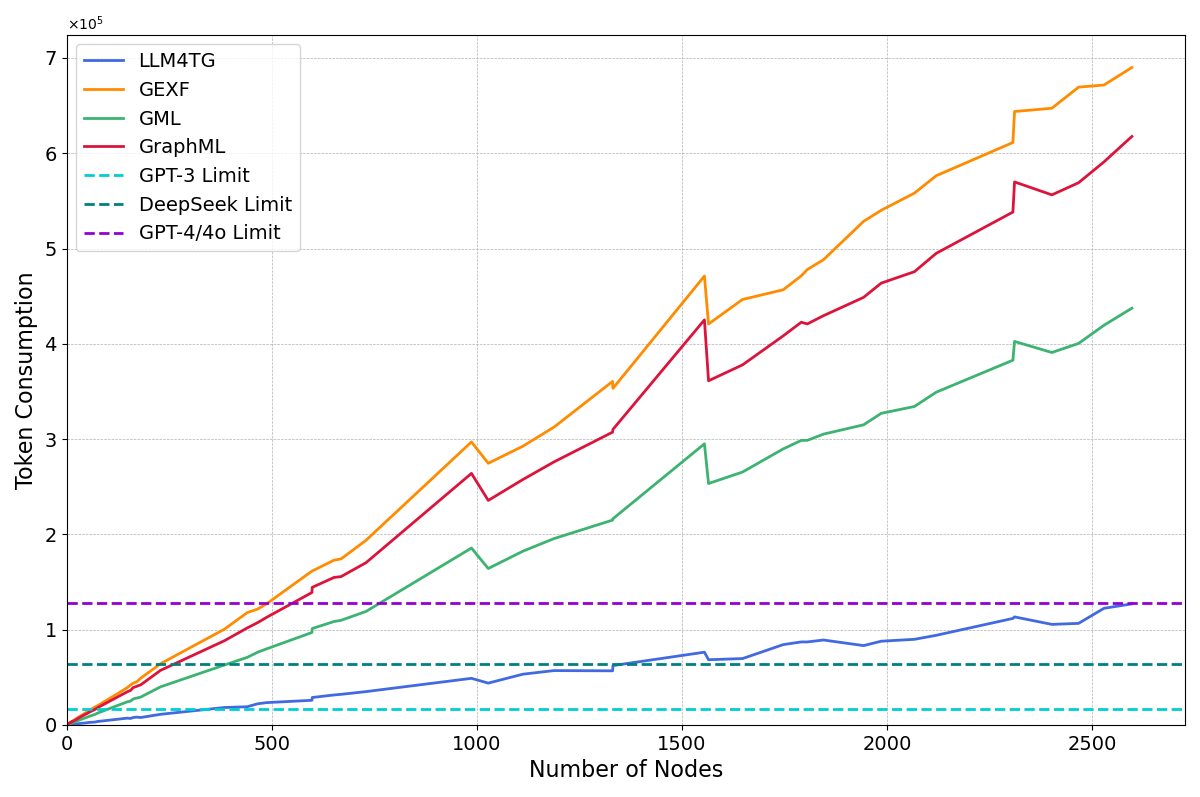}
    \caption{Token consumption in different graph formats}
    \vspace{-1\baselineskip}
    \label{fig:tokens}
\end{figure}

\smallskip
\noindent\textbf{Token Consumption.}
Fig.~\ref{fig:tokens} compares token consumption using the \texttt{cl100k\_base} tokenizer. LLM4TG substantially reduces token usage compared with GEXF, GML, and GraphML, whose syntactic overhead limits the graph sizes that can fit within LLM context windows. Token consumption also depends on edge counts and attributes, explaining variations among graphs of similar sizes. By reducing representation overhead, LLM4TG allows more transaction information to be included within a fixed token budget, while CETraS further reduces graph sizes for few-shot tasks.

\subsection{Capabilities and Limitations}

Our findings reveal substantial differences across the three levels of transaction graph understanding. LLMs accurately retrieve explicit node attributes and identify meaningful graph characteristics, but perform less reliably on global metrics requiring numerical comparisons. At the contextual interpretation level, feature-based classification achieves 39.83\%--46.68\% top-1 accuracy, while GPT-4o reaches 50.49\% using raw graphs. Although top-3 accuracy reaches 72.43\%, specialized machine learning models generally remain more effective for precise classification.

Several factors constrain LLM-based transaction analysis. Token limits restrict the amount of graph information and the number of few-shot references, making reference selection and graph sampling important. Engineered features provide compact inputs, whereas raw graphs retain richer structural information at higher token costs. Furthermore, LLM-generated explanations may contain inaccurate interpretations even when predictions are correct. These findings support the use of LLMs for exploratory transaction analysis while highlighting the need for more reliable graph reasoning and explanation verification.

%=========================================================
\section{Conclusion} 
%=========================================================

This work systematically evaluates LLMs for Bitcoin transaction graph analysis through a three-level framework. We introduce LLM4TG and CETraS to reduce token consumption while preserving transaction-relevant information and graph connectivity. Experimental results show that LLMs effectively extract node-level information and identify meaningful graph characteristics, but struggle with global graph reasoning and precise address classification. Comparisons between raw graphs and engineered features reveal substantial differences across models, with raw graphs achieving higher classification accuracy for GPT-4o at greater token costs. These findings highlight the potential of LLMs for exploratory cryptocurrency transaction analysis and identify key challenges in classification accuracy and explanation reliability.

\section*{Acknowledgment}
This research was supported by the OpenAI Researcher Access Program (Project ID: 0000007730).

\bibliographystyle{IEEEtran}  
\bibliography{bib}

\end{document}